\documentclass[12pt]{article}

\newcommand{\be}{\begin{equation}} \newcommand{\ee}{\end{equation}}
\newcommand{\bea}{\begin{eqnarray}} \newcommand{\eea}{\end{eqnarray}}
\newcommand{\el}{\nonumber \\}

\newcommand{\adot}{\dot{a}} \newcommand{\addot}{\ddot{a}}
\newcommand{\bdot}{\dot{b}} \newcommand{\bddot}{\ddot{b}}
\newcommand{\ndot}{\dot{n}}

\newcommand{\phidot}{\dot{\phi}}
\newcommand{\phiddot}{\ddot{\phi}}

\def\am{Center For Theoretical Physics, Department of Physics\\ Texas A\&M University,
 College Station,TX 77843-4242,
USA}
\def\regina{**Department of Physics, University of Regina\\ Regina SK, S4S OA2, Canada}
\def\address#1{\begin{center}{ \it #1} \end{center}}
\def\author#1{\begin{center}{ \sc #1} \end{center}}
\def\title#1{\begin{center} {\Large #1 } \end{center}}
\def\Journal#1#2#3#4{{#1} {\bf #2}, #3 (#4)}

\def\EJC{{\em Eur.Phys.J} C}
\def\JHEP{\em JHEP}

\def\NPB{{\em Nucl. Phys.} B}
\def\PLB{{\em Phys. Lett.}  B}
\def\PRL{\em Phys. Rev. Lett.}
\def\PRD{{\em Phys. Rev.} D}

\begin{document}

\begin{titlepage}

\title{Five Dimensional Cosmology in Horava-Witten M-Theory}
\author{R. Arnowitt, James Dent, and B. Dutta**}
\address{\am}
\address{\regina}

\begin{abstract}
The cosmology in the Hubble expansion era of the Horava-Witten M-theory compactified on a Calabi-Yau threefold is studied in the reduction to five-dimensions where the effects of the Calabi-Yau manifold are summarized by the volume modulus, and all perturbative potentials are included.  Matter on the branes are treated as first order perturbations of the static vacuum solution, and all equations in the bulk and all boundary conditions on both end branes are imposed.  It is found that for a static volume modulus and a static fifth dimension, y, one can recover the four dimensional Robertson Friedmann Walker cosmology for relativistic matter on the branes, but not for non-relativistic matter.  In this case, the Hubble parameter $\mathnormal{H}$ becomes independent of y to first order in matter density.  This result holds also when an arbitrary number of 5-branes are included in the bulk.  The five dimensional Horava-Witten model is compared with the Randall Sundrum phenomenology with a scalar field in the bulk where a bulk and brane potential are used so that the vacuum solutions can be rigorously obtained.(In the Appendix, the difficulty of obtaining approximate vacuum solutions for other potentials is discussed.)  In this case non-relativistic matter is accommodated by allowing the distance between the branes to vary.  It is suggested that non-perturbative potentials for the vacuum solution of Horava-Witten theory are needed to remove the inconsistency that non-relativistic matter creates.
\end{abstract}

\end{titlepage}

\section{Introduction}
Over the past several years there has been much interest in examining cosmology in five dimensions to see if the conventional four-dimensional Robertson Walker Friedman(RWF) cosmology can be recovered.  Theoretical motivation for such attempts can be found within the framework of the Horava-Witten M-theory \cite{hw,hw2,witten,horava}, the strong coupling limit of the heterotic string theory.  In Horava-Witten (HW) theory, space is eleven-dimensional (11D) bounded by two ten-dimensional orbifold planes with a $Z_2$ reflection symmetry in the eleventh dimension.  In the lowest approximation $M_{10}$ = $M_4$$\times$X$\times$$S^1$/$Z_2$ where X is a compact Calabi-Yau (C-Y) threefold.  Eleven dimensional supergravity resides in the bulk.  The construction of a consistent theory depends on a set of interlocking constraints due to anomaly cancelation, gauge invariance, and local supersymmetry invariance.  Thus there must be $E_8$ gauge interactions with chiral multiplets on each orbifold plane (one being the physical world, the other the hidden sector) and a consistent theory exists only if their exists a relation between the ten-dimensional gauge coupling constant and the eleven-dimensional Planck mass.  Assuming the compactification mass is the GUT mass, $M_G$, this latter relation then shows that the eleventh dimension is $\cal{O}$(10) times larger than the C-Y scale.  Dimensional reduction then leads naturally to a five dimensional world $M_4$$\times$$S^1$/$Z_2$, with the residual effects of the C-Y manifold being described by their moduli.  The details of this formulation have been extensively developed in \cite{ovrut}.

Phenomenological analysis of five-dimensional cosmology was stimulated by the work of Binetruy, Deffayet, and Langlois \cite{binetruy}, and subsequently by the Randall-Sundrum (RS) model \cite{rs,rs2}.  In the RS1 model, where space is five dimensional with five-dimensional gravity in the bulk and physical matter on two four dimensional bounding planes, it has been argued that the standard four-dimensional RFW cosmology is obtained only if the fifth dimension is stabilized \cite{csaki} (though counter arguments have been given in $\cite{khoury}$).  In order to achieve this naturally, it has been suggested that one phenomenologically add a scalar field in the bulk with appropriate bulk and brane potentials \cite{gw,dewolfe}.  Vacuum solutions appropriate for these potentials are then obtained and solutions are discussed with brane matter viewed as perturbations on the vacuum\cite{cline,cline2}.

In this paper we examine the five-dimensional reduction of HW theory in the simplest approximation of keeping only the volume moduli of the C-Y space, and consider under what circumstances one can reproduce the standard RWF cosmology.  When no matter is present on the branes, a vacuum solution of the gravitational equations has already been given in \cite{ovrut}.  This vacuum solution automatically fixes the bulk and brane cosmological constants (without any fine tuning) so that when matter is added to the branes the net cosmological constant is correctly zero.  Treating the matter as a pertubation on the vacuum, we find that the standard cosmology can be recovered for relativistic matter for a static solution but not for non-relativistic matter.  We further show that this result remains even when an arbitrary number of five-branes are added in the bulk.  We then examine the RS phenomenology, which closely resembles the 5D reduction of HW M-theory.  Vacuum solutions for this system are difficult to find for an arbitrarily chosen bulk potential. Further, the difficulty arising from non-relativistic matter in the HW theory is evident only when the boundary conditions on both orbifold planes are rigorously imposed.  Therefore, we examine the class of models presented in \cite{dewolfe} which can produce exact vacuum solutions for a specific type of potential.  We show that the inconsistencies that arise in the HW theory due to the presence of non-relativistic matter can be avoided due to the introduction of ad hoc potentials not found in the HW M-theory with a subsequent fine tuning needed to set the brane cosmological constant to zero.(In contrast, HW theory sets the cosmological constant to zero without any fine tuning.)  The solution is then static only to lowest order, the brane separations varying in time due to the presence of non-relativistic matter.  In an appendix we examine the RS model with the bulk potential given in \cite{cline}, and discuss the difficulties in finding the vacuum metric for this case.  

In Section 2 we briefly review the Horava-Witten model and its 5D reduction, and write down the field equations and boundary conditions.  In Section 3 these equations are solved in the presence of matter, and we show that non-relativistic matter is excluded unless one fine tunes the matter on the two branes.  In Section 4 we introduce 5-branes in the bulk (the only additional freedom allowed in HW theory) which reduce to 3-branes in the 5D reduction, and show the same problem remains (unless the non-relativistic matter on the end branes and on the bulk 5-branes is fine tuned).  In Section 5 we relax the static condition imposed in earlier sections and find that the fine-tuning of the matter on the branes might be alleviated due to the non-trivial bulk dependence of the time derivative of the scalar field.  In Section 6 we discuss the RS1 model and show how it differs from HW theory.  Conclusions are given in Section 7.  An appendix discusses the choice of potentials in RS1 used in \cite{cline}.  It is shown that the approximate solution for the vacuum metric used there does not give rise to a hierarchy when all the boundary conditions on both branes are imposed.

Previous work on HW cosmology has been given in \cite{enq,ellwanger,lukas}.  However, the first two papers do not impose all the boundary conditions, and hence do not see the difficulties found here.  The last paper is concerned with the inflationary era rather that the Hubble expansion era being discussed here.  A very general analysis for an arbitrary model was given in \cite{kobayashi}, but the authors do not seem to have noticed the difficulty discussed here.  Within the M-theory framework, there have been several papers suggesting that stabilization of moduli can be achieved by turning on fluxes \cite{buchbinder,linde,becker}.  The first two give rise to large negative cosmological constants, while the last to a large positive cosmological constant.  How to reduce these constants to their physics values in a natural way remains (as well as how to analyse the presence of matter).

\section{Horava-Witten M-theory}
In this section we review the basic formulae of Horava-Witten M-theory.
Horava and Witten \cite{hw,hw2} showed that the low energy limit of eleven dimensional supergravity on the orbifold $R^{10}$$\times$$S^1$/$Z_2$
is dual to the strong coupling limit of the ten dimensional $E_8$$\times$$E_8$ heterotic string.  This theory is reduced to five dimensions via compactification on a
Calabi-Yau threefold of volume $\mathcal{V}$.  In order to cancel the
gauge and gravitational anomalies that arise,  an $E_8$ gauge group is required to reside on each of the two 10D planes at the
orbifold fixed points ($x^{11}$=0 and $\pi$$\rho$), and a relation between the Yang-Mills gauge coupling constant and
the eleven-dimensional gravitational coupling constant is produced \cite{witten}.  This is remarkable in that it explains the
discrepancy between the GUT scale  $M_G$$\simeq$3$\times$$10^{16}$GeV and the four-dimensional Planck mass, since it is the
eleven-dimensional Planck mass, which is $\simeq$ $M_G$, that is the fundamental scale.  Upon reduction to four dimensions
and using the values of the four-dimensional Newton constant $G_N$, as well as $\alpha_{GUT}$ and $M_G$, one
finds that the orbifold radius is about ten times the compactification scale, rendering the theory effectively
five-dimensional.  

Discarding the shape moduli and keeping only the volume modulus, $\mathnormal{V}$, of the C-Y threefold, the bosonic part of the reduced five-dimensional Lagrangian takes the
following form \cite{ovrut}
\begin{eqnarray}\nonumber
S & = &-\frac{1}{2\kappa_{5}^2}\int_{M_5}\sqrt{g}[R
+\frac{1}{2}V^{-2}\partial_{\alpha}V\partial^{\alpha}V+\frac{3}{2}\alpha^{2}V^{-2}]\\\nonumber& &+\frac{1}{\kappa_{5}^2}\sum_{i}\int_{M_{4}^{(i)}}\sqrt{-g}V^{-1}3(-1)^{i+1}\alpha
\\\nonumber& &-\frac{1}{16\pi\alpha_{GUT}}\sum_{i}\int_{M_{4}^{(i)}}\sqrt{-g}VtrF_{\mu\nu}^{i^{2}}
\\& &
-\sum_{i}\int_{M_{4}^{(i)}}\sqrt{-g}\bigg[(D_{\mu}C)^{n}(D_{\mu}\bar{C})^{n}+V^{-1}\frac{\partial{W}}{\partial{C^{n}}}\frac{\partial{\bar{W}}}{\partial{\bar{C^{n}}}}
+D^{(\mu)}D^{(\mu)}\bigg]
\end{eqnarray}
where $\kappa_5$ is the five-dimensional Newton constant,  R is the
five-dimensional Ricci scalar, $\alpha_G$ is the GUT scale coupling, $F_{\mu\nu}^{(i)}$ are the gauge field strengths on the boundary orbifolds, 
$C^n$ are complex scalars of chiral matter, W is the superpotential, and the last term represents the D term of the gauge theory on the branes.  The parameter $\alpha$, which is $\cal{O}$($10^{15}$GeV), fixes the bulk and brane cosmological
 constants.  It is defined as \cite{ovrut2}
\begin{displaymath}
\alpha = -\frac{1}{8\sqrt{2}\pi\mathcal{V}}\left(\frac{\kappa}{4\pi}\right)^{2/3}\int_{X}\omega\wedge trR\wedge R\nonumber
\end{displaymath}
where R is the Calabi-Yau curvature constructed from the metric $g_{a\bar{b}}$, and $\omega_{a\bar{b}}$ is the K$\ddot{a}$hler form
\begin{equation}
\omega_{a\bar{b}} = ig_{a\bar{b}}
\end{equation}  
with holomorphic and anti-holomorphic indices a and $\bar{b}$.

From this action and the ansatz
\begin{equation}
ds^{2}=a(t,y)^{2}dx^{k}dx^{k}-n(t,y)^{2}dt^{2}+b(t,y)^{2}dy^{2},
\end{equation}
where y$\equiv$$x^{11}$ is the coordinate of the fifth-dimension which extends from y=0 to y=$\pi$$\rho$, and t is time,  
one can obtain the five-dimensional Einstein field equations
\begin{eqnarray} 
  G^t_{\ t} &=& \frac{3}{b^2} \left[\frac{a''}{a}+\frac{a'}{a}\left(\frac{a'}{a}-\frac{b'}{b}\right)\right] -\frac{3}{n^2}\frac{\adot}{a}\left(\frac{\adot}{a}+\frac{\bdot}{b}\right) \el
  &=& -\frac{1}{4} n^{-2}\phidot^2 - \frac{1}{4} b^{-2}\phi'{}^2 - \frac{3}{4}\alpha^2 e^{-2\phi} \\ & &-
  \frac{1}{M_5^3}\sum_{i=1}^2 \delta(y-y_i) b^{-1}\left(\rho_{ir} e^{\phi_i} + \rho_{inr} e^{-\phi_i}+ 3
  M_5^3\alpha_i e^{-\phi_i}\right) \el
  G^k_{\ k} &=& \frac{1}{b^2} \left[2\frac{a''}{a}+\frac{n''}{n}+\frac{a'}{a}
\left(\frac{a'}{a}+2\frac{n'}{n}\right)-\frac{b'}{b}\left(\frac{n'}{n}
+2\frac{a'}{a} \right) \right] \el
  &&-\frac{1}{n^2}\left[2\frac{\addot}{a}+\frac{\bddot}{b}+\frac{\adot}{a}
\left(\frac{\adot}{a}-2\frac{\ndot}{n}\right)
+\frac{\bdot}{b}\left(2\frac{\adot}{a}-\frac{\ndot}{n}\right)\right] \el
  &=& \frac{1}{4} n^{-2}\phidot^2-\frac{1}{4} b^{-2}\phi'{}^2-\frac{3}{4}\alpha^2 e^{-2\phi}  \\ & & +
  \frac{1}{M_5^3}\sum_{i=1}^2\delta(y-y_i) b^{-1} \left(p_{ir} e^{\phi_i} - 3 M_5^3\alpha_i e^{-\phi_i}\right) \el
  G^y_{\ y} &=& \frac{3}{b^2} \frac{a'}{a} \left( \frac{a'}{a} +\frac{n'}{n}\right) - \frac{3}{n^2}\left[\frac{\addot}{a} + \frac{\adot}{a} \left(
\frac{\adot}{a} -\frac{\ndot}{n} \right)\right] \el
  &=& \frac{1}{4} n^{-2}\phidot^2+\frac{1}{4} b^{-2}\phi'{}^2-\frac{3}{4}\alpha^2 e^{-2\phi}\\
  G_{ty} &=& 3\left( \frac{n'}{n}\frac{\adot}{a}+\frac{a'}{a}\frac{\bdot}{b}
-\frac{\adot'}{a} \right) = \frac{1}{2}\phidot\, \phi'  \ ,
\end{eqnarray}
where i = 1, 2 corresponds to the fixed points at y=0, $\pi$$\rho$, prime and dot denote derivatives with
respect to y and t respectively, and V = $e^{\phi}$ where
$\phi$ is the breathing modulus of the Calabi-Yau.   The non-relativistic matter density on the i'th
orbifold is $\rho_{inr}$, the relativistic matter is $\rho_{ir}$, and $p_{ir}$ is the pressure.  

The $\delta$-functions in Eqs.(4) and (5) imply boundary conditions at the orbifolds y=0, $\pi$$\rho$ given by

\begin{eqnarray}
(-1)^{i}\frac{1}{b}\frac{a'}{a} \,\bigg|_{y=y_{i}}&=& \frac{\rho_{i}}{6M_{5}^{3}}\,\,\,;\,\,\,\rho_i = \rho_{ir} e^{\phi_i} + \rho_{inr} e^{-\phi_i}+ 3
  M_5^3\alpha_i e^{-\phi_i}\\
(-1)^{i}\frac{1}{b}\frac{n'}{n} \,\bigg|_{y=y_{i}}&=& -\frac{2\rho_{i}+3p_{i}}{6M_{5}^{3}}\,\,\,;\,\,\,p_i = p_{ir} e^{\phi_i} - 3 M_5^3\alpha_i e^{-\phi_i}
\end{eqnarray}
\begin{equation}
\alpha_i = (-1)^i\alpha
\end{equation}
where $M_5$ is the five-dimensional Planck mass and $\rho_{i}$ and $p_{i}$ are the total matter density and pressure on the two orbifolds.  Thus the bulk cosmological constant $\alpha$, and the brane cosmological constant $\alpha_i$ are naturally correlated without any fine tuning.  

In addition to the Einstein field equations, one can derive field equations and boundary conditions
for the breathing modulus from the action:
\bea 
  - n^{-2} \left[\phiddot + \left(-\frac{\ndot}{n}+3\frac{\adot}{a}+\frac{\bdot}{b}\right)\phidot\right] \el
  + b^{-2} \left[\phi'' + \left(\frac{n'}{n}+3\frac{a'}{a}-\frac{b'}{b}\right)\phi'\right] + 3\,\alpha^2 e^{-2\phi} &=& 0 \\
   \left(\phi'-(3b\alpha - \frac{b}{M^{3}}\rho_{inr})e^{-\phi}\right)\,\bigg|_{y=y_{i}}&=& 0.
\eea
This differs from the result of Ref.\cite{enq} in that non-relativistic matter has been included in the boundary condition for
$\phi$ as it should since $\phi$ is coupled to the gauge fields on the branes at y=0 and $\pi$$\rho$ by the factor $\mathnormal{V^{-1}}$=$e^{-\phi}$ in the superpotential term in Eq.(1). (Note that $\phi$ also couples to gauge fields in Eq.(1) with the factor $\mathnormal{V}$=$e^{\phi}$, but the coefficient $F_{\mu\nu}$$F^{\mu\nu}$ vanishes for the radiation fields).

\section{Solution of the 5D equations}
We now proceed to solve the field equations using a perturbative expansion in powers of 
matter on the branes.  This expansion is allowed since in the Hubble era the matter density is very small
compared to the cosmological constants in the bulk and on the brane which are of GUT size.  We start with the
vacuum solutions and then include matter on the branes as a higher order correction.  The vacuum solution
which fully solves the bulk and boundary equations, preserves Poincaire invariance, and breaks 4 of the 8
supersymmetries (appropriate for getting N=1 supergravity when one descends to four dimensions) was given in $\cite{ovrut}$ as

\begin{equation}
  a(y)= f^{\frac{1}{2}} \,; \,\, n(y)= f^{\frac{1}{2}} \,;\,\, b(y)=b_of^2 \,;\,\,V(y)=b_of^3.
\end{equation}
Here $b_o$ is a constant that is arbitrary due to the flat directions of the potential and f is given by
\begin{equation}
  f(y)=c+\alpha\big| y \big|.
\end{equation}
where c is a constant.
To first order in $\rho$, the vacuum solutions are perturbed to take the following form:
\begin{eqnarray}
a(y,t) &= &f^{1/2}(1+\delta a(y,t))\\
n(y) &= &f^{1/2}(1+\delta n(y))\\
b(y) &= &b_{0}f^{2}(1+\delta b(y,t))\\
V(y) &= &b_{0}f^{3}(1+\delta V(y,t))
\end{eqnarray}
It is convenient to introduce the notation
\begin{eqnarray}
\Delta a' \equiv \delta a' + \frac{\alpha}{2f}\delta V - \frac{\alpha}{2f}\delta b\\
\Delta n' \equiv \delta n' + \frac{\alpha}{2f}\delta V - \frac{\alpha}{2f}\delta b\\
\Delta V' \equiv \delta V' + \frac{3\alpha}{f}\delta V - \frac{3\alpha}{f}\delta b
\end{eqnarray}
along with the definition of the Hubble constant
\begin{equation}
H \equiv \frac{\dot{a}}{a}.
\end{equation}
The significance of the combinations of Eqs.(19-21) is that they are invariant under a first order coordinate change in the y coordinate: $\bar{y}$ = y + $\delta$(y).

To first order in $\rho$, the Einstein equations $G_{tt}$, $G_{kk}$, $G_{yy}$, and the field equation for the breathing modulus
become the following:
\begin{eqnarray}
\Delta a'' = b_{o}^2 f^3\left(H^2+H\frac{\dot{b}}{b} - \frac{1}{12}\dot{\phi}^2\right)\equiv b_{o}^2
f^3A_1 \\
\Delta n'' +2\Delta a'' = b_{o}^2 f^3\left(3H^2+2\dot{H}+2H\frac{\dot{b}}{b} + \frac{\ddot{b}}{b}+
\frac{\dot{\phi}^2}{4}\right)\equiv b_{o}^2 f^3A_2\\
3\Delta a' +\Delta n' -\Delta V' = \frac{b_{o}^2 f^4}{\alpha}\left(4H^2+2\dot{H} -
\frac{\dot{\phi}^2}{6}\right)\equiv \frac{b_{o}^2 f^4}{\alpha}A_3\\
\Delta V'' +\frac{3\alpha}{f}\left(\Delta n' +3\Delta a' - \Delta V' \right)
=b_{o}^2f^3\left(\ddot{\phi}+3H\dot{\phi}+\frac{\dot{b}}{b}\dot{\phi}\right)
\equiv b_{o}^2f^3A_4
\end{eqnarray}
As we will show in Section 5, the $G_{ty}$ equation is of higher order and will be discussed later concerning the possibility of
including y-dependence in the Hubble constant.  Inserting the metric ansatz into the boundary equations (8),(9), and (12)
yields
\begin{eqnarray}
\Delta a_i' = (-1)^i\left(\frac{b_{o}^2f_{i}^5}{6M_{5}^{3}}\rho_{ir}+\frac{1}{6M_{5}^{3}f_i}\rho_{inr}\right)\\
\Delta n_i' = (-1)^{i+1}\left(\frac{b_{o}^2f_{i}^5}{6M_{5}^{3}}\left(2\rho_{ir}+3p_{ir}\right)+\frac{2}{6M_{5}^{3}f_i}\rho_{inr}\right)\\
\Delta V_i' = (-1)^{i+1}\frac{1}{M_{5}^{3}f_i}\rho_{inr}
\end{eqnarray}
where $f_i$ are the values of f(y) on the branes:
\begin{equation}
f_1 \equiv c\,\,\,;\,\,\,f_2 \equiv c + \alpha_2\pi\rho
\end{equation}
It will also be helpful to use the combination of Eq.(27) and Eq.(28) that isolates $\rho_{inr}$:
\begin{equation}
\left(3\Delta a' + \Delta n'\right)_{y_i} = (-1)^{i}\frac{1}{6M_{5}^{3}f_i}\rho_{inr}
\end{equation}
where we have used $p_{ir}$=$\rho_{ir}$/3.  The significance of the above results is that $\emph{both}$ the field equations $\emph{and}$ the
boundary conditions can be expressed in terms of the y-invariant combinations of Eqs.(19-21).

We are now ready to examine the solution of the bulk equations, impose the boundary conditions on them, and check
their consistency.  First of all we notice that the equations $G_{tt}$ and $G_{kk}$ are easily solved by integration
with respect to y.  However, it will be more convenient to change integration variables from y to f and to use the combination 
$G_{tt}$ + $G_{kk}$ instead of $G_{kk}$.  One finds
\begin{eqnarray}
\Delta a' = \frac{b_{o}^2}{\alpha}\left( \int_{f_1}^{f}df'f'^{3} A_1 + c_1\right)\\
3\Delta a' + \Delta n' = \frac{b_{o}^2}{\alpha}\left( \int_{f_1}^{f}df'f'^{3}\left(A_1+A_2\right) + c_1 + c_2\right)
\end{eqnarray} 
and imposing the boundary conditions Eqs.(27) and (28) one obtains
\begin{eqnarray}
c_1 = -\frac{\lambda}{6}\left(b_{o}^2f_{1}^5\rho_{1r}+\frac{1}{f_1}\rho_{1nr}\right)\,\,\,\, ; \,\,\,\,\lambda \equiv
\frac{\alpha}{b_{o}^2 M_{5}^3}\\
\int_{f_1}^{f_2}df'f'^3A_1 = \frac{\lambda}{6}\left(b_{o}^2\left(\rho_{1r}f_{1}^5+\rho_{2r}f_{2}^5\right) + \left(\frac{\rho_{1nr}}{f_1}+\frac{\rho_{2nr}}{f_2}\right)\right)\\
c_1 + c_2 = -\frac{\lambda}{6f_1}\rho_{1nr}\\
\int_{f_1}^{f_2}df'f'^{3}\left(A_1+A_2\right) =
\frac{\lambda}{6}\left(\frac{1}{f_1}\rho_{1nr}+\frac{1}{f_2}\rho_{2nr}\right).
\end{eqnarray}

Eqs.(35) and (37), arising from the boundary conditions at the distant brane y=$\pi$$\rho$, thus produce constraints on the time derivatives (which enter in the $A_i$ defined in Eqs.(23-26)) of the metric and $\phi$.  We next integrate the field equation for the breathing modulus, Eq.(26), using Eq.(25) to eliminate 3$\Delta a'$ + $\Delta n'$ -$\Delta V'$.  Combined with the boundary condition Eq.(29) one obtains
\begin{equation}
\Delta V' = \frac{b_{o}^2}{\alpha}\left( \int_{f_1}^{f}df'f'^{3}\left(A_4-3A_3\right) + c_3\right)\,\,\,;\,\,\,c_3 =
\frac{\lambda}{f_1}\rho_{1nr}
\end{equation} 
\begin{equation}
\int_{f_1}^{f_2}df'f'^{3}\left(A_4-3A_3\right) = -\lambda\left(\frac{1}{f_1}\rho_{1nr}+\frac{1}{f_2}\rho_{2nr}\right).
\end{equation}
The remaining equation to be satisfied is Eq.(25).  Inserting Eqs.(33) and (38) back into Eq.(25) gives the constraint
\begin{equation}
\frac{b_{o}^2}{\alpha}\left(\int^{f}_{f_1}df'f'^3\left(A_1+A_2+3A_3-A_4\right) +c_1+c_2-c_3\right) = \frac{b_{o}^2f^4}{\alpha}A_3
\end{equation}
This constraint is a strong one as it must hold for all y.

We now examine the consistency of this system.  We consider here the static case where $\dot{\phi}$ = 0 = $\dot{b}$.  Here $A_4$=0 and $A_3$ = $A_1$+$A_2$.  Thus multiplying Eq.(37) by 3 and adding to Eq.(39) gives
\begin{equation}
0 = -\frac{\lambda}{2}\left(\frac{1}{f_1}\rho_{1nr}+\frac{1}{f_2}\rho_{2nr}\right)
\end{equation}
Thus a consistent solution without fine tuning requires (when $\dot{\phi}$ = 0 = $\dot{b}$)
\begin{equation}
\rho_{nr} = 0
\end{equation}
i.e. only relativistic matter is consistent with Horava-Witten cosmological equations.  However, in addition to Eq.(42) one must also make sure that the constraint Eq.(40) is satisfied.    In Section 5 we will show that the $G_{ty}$ equation implies $\mathnormal{H'}$ is $\mathcal{O}$($\rho^{3/2}$) for the static case, and hence to $\mathcal{O}$($\rho$) that we are calculating one can consider $\mathnormal{H^2}$ and $\dot{H}$ to be independent of y.  Hence we note that for the static case Eqs.(35) and (37) correctly reduce to the RFW cosmology equations for relativistic matter with $G_N$ defined in terms of $\lambda$ and $f_i$:
\begin{equation}
H^2 = \frac{8\pi}{3}G_N\left(\rho_{1r}' + \rho_{2r}'\right)\,\,;\,\,G_N = \frac{\lambda}{4\pi}\left(\frac{1}{f_{2}^4-f_{1}^4}\right)\end{equation}
\begin{equation}
H^2 + 2\dot{H} = 0
\end{equation}
(where the rescaled $\mathnormal{\rho_{ir}'}$ = $\mathnormal{b_{o}f_{i}^5\rho_{ir}}$ are the mass densities as seen locally on the orbifold 3-branes).  Eqs.(43) and (44) just incorporate the 4D relativistic matter equation of continuity: $\dot{\rho}$ = -3$\mathnormal{H}$($\rho$ + p) = -4 $\mathnormal{H\rho}$.  Since $A_1$ + $A_2$ = $A_3$,  Eq.(40) is then identically satisfied as a consequence of Eqs.(42) and (44).

In summary we note that it is the boundary conditions on the distant brane at y = $\pi$$\rho$, Eqs.(37) and (39), that produces the constraint Eq.(42) on non-relativistic matter (which is why earlier analyses have not seen this).  However, a satisfactory FRW cosmology does result for relativistic matter, with the brane cosmological constant naturally vanishing with no fine tuning required.

\section{Inclusion of 5-Branes}
We have shown that in the static case ($\dot{\phi}$, $\dot{b}$ = 0) the system of bulk equations with their boundary conditions imposed is inconsistent when
non-relativistic matter is included in the system.  Therefore we would like to examine other situations that
might lead to a consistent solution when all types of matter are present.  The only additional generalization available in the Horava-Witten theory is to include a set of 5-branes in the bulk transverse to the orbifold direction \cite{witten,strominger}.  We follow here the analysis of $\cite{brandle}$ of a single 5-brane residing at an arbitrary position y=Y in the bulk (which can easily be generalized to an arbitrary number of 5-branes).  The fields that live on the 5-brane include an N=1 chiral multiplet and N=1 gauge multiplets but no
superpotential. 
One must generalize the function f(y) and the definition of $\alpha$ to be (for 0$\leq$y$\leq$$\pi$$\rho$)
\begin{eqnarray}
f(y) = c + h(y)\,\,;\,\, h(y) = -\alpha_1y+\left(-\alpha_5y+\alpha_5Y\right)\theta\left(y-Y\right)\\
h'(y) \equiv \alpha(y) = -\alpha_1 -\alpha_5\theta\left(y-Y\right)\\
 \alpha(y=Y)\equiv\alpha_3= \frac{\alpha_5}{2}
\end{eqnarray}
with the cohomology condition
\begin{equation}
\alpha_1+\alpha_2+\alpha_5 = 0
\end{equation}
but otherwise the vacuum solution has the same form as Eq.(13).  Note that f(y) is continuous whereas $\alpha$(y) = $h'$(y) is not
\begin{equation}
\alpha(y) = \left\{ \begin{array}{ll} -\alpha_1 & 0\leq y < Y\\-\alpha_1 - \alpha_5 = \alpha_2 & Y < y \leq{\pi\rho}\end{array}\right.
\end{equation}

We use the same ansatz as before for the metric: Eqs. (14),(15),(16),(17), and (18).  However, the Einstein equations are
altered due to the fact that $\alpha$ is no longer a constant; thus $\alpha'$ no longer vanishes.  We also must
make some assumptions about the matter content of the five-brane in the bulk.  We know that in order to give rise to the
Big Bang at the end of inflation that the inflaton couples to matter on the physical orbifold at y=0 and we have assumed that
it also couples to matter on the orbifold at y=$\pi$$\rho$.  There is no a priori reason to believe that it also couples to
any matter fields on the 5-brane in the bulk.  However, we will make the assumption that it does couple to the
five-brane matter with the same strength V as for the two orbifolds (this does not effect the general conclusions of this section).  

Now we would like to solve the Einstein equations in the presence of this 5-brane.  The calculation is very similar
to that done in the previous section with the modifications that i runs from 1 to 3 and $\alpha'$ terms must now be
included.  For example in the $G_{tt}$ equation we still have the definition
\begin{equation}
\Delta a' \equiv \delta a' + \frac{\alpha}{2f}\delta V - \frac{\alpha}{2f}\delta b.\\
\end{equation}
However $\Delta a^{''}$ is
\begin{equation}
\Delta a'' = \delta a'' +\frac{\alpha^2}{2f^2}\left(\delta b - \delta V\right) - \frac{\alpha}{2f}\left(\delta b' -
\delta V'\right) -\frac{\alpha'}{2f'}\left(\delta b - \delta V\right)
\end{equation}
and after making use of the ansatz, the $G_{tt}$ equation becomes
\begin{equation}
\Delta a'' = -\frac{1}{M_{5}^3}\left(\rho_{3r}b_{o}^2 f_{3}^5 +\frac{1}{f_3}\rho_{3nr}\right)\delta\left(y-Y\right)
+b_{o}^2 f^3 A_1
\end{equation}
where the $\delta$-function term arises from a y-derivative of $\alpha$, and $A_1$ is given in Eq.(23).  The key point
in this relation is the unexpected result that the $\alpha_5$ terms cancel, not only at the vacuum order (which was already verified in showing Eq.(13) with f(y) of Eq.(45) is correctly the vacuum solution), but also at higher orders.
This also applies to the other field equations. 

Since there is a discontinuity in $\alpha$ when crossing the five-brane, we solve the $G_{tt}$ equation in two domains: 0$\leq$y$<$Y and Y$<$y$\leq$$\pi$$\rho$ and
then match them across y$=$Y.  We define the gauge covariant combinations in these regions as
\begin{eqnarray}
\Delta a_{1}' \equiv \delta a' +\frac{\alpha_1}{2f}\left(\delta b - \delta V\right)\,\,;\,\,0\leq y <Y\\
\Delta a_{2}' \equiv \delta a' -\frac{\alpha_2}{2f}\left(\delta b - \delta V\right)\,\,;\,\,Y<y\leq\pi\rho.
\end{eqnarray}
These quantities obey the boundary conditions
\begin{eqnarray}
\Delta a_{1}'(y=0) = -\frac{1}{M_{5}^3}\left(\rho_{1r}b_{o}^2 f_{1}^5 +\frac{1}{f_1}\rho_{1nr}\right)\\
\Delta a_{2}'(y=\pi\rho) = \frac{1}{M_{5}^3}\left(\rho_{2r}b_{o}^2 f_{2}^5 +\frac{1}{f_2}\rho_{2nr}\right).
\end{eqnarray}
We now solve the $G_{tt}$ equation in these two domains to obtain
\begin{eqnarray}
\Delta a_{1}' = \int_{0}^{y}dy'b_{o}^2 f^3A_1-\frac{1}{M_{5}^3}\left(\rho_{1r}b_{o}^2 f_{1}^5
+\frac{1}{f_1}\rho_{1nr}\right)\\
\Delta a_{2}' = -\int_{y}^{\pi\rho}dy'b_{o}^2 f^3A_1+\frac{1}{M_{5}^3}\left(\rho_{2r}b_{o}^2 f_{2}^5
+\frac{1}{f_2}\rho_{2nr}\right).
\end{eqnarray}
The discontinuity across the five-brane required by Eq.(52) gives
\begin{equation}
\int_{Y-\epsilon}^{Y+\epsilon}dy'\Delta a'' = \left(\Delta a_{2}'-\Delta a_{1}'\right)_{y=Y}=
-\frac{1}{M_{5}^3}\left(\rho_{3r}b_{o}^2 f_{3}^5 +\frac{1}{f_3}\rho_{3nr}\right)
\end{equation}
Thus, subtracting (57) from (58) leads to
\begin{equation}
\int_{0}^{\pi\rho}dy'b_{o}^2 f^3A_1 = \sum_{i=1}^3\frac{1}{M_{5}^3}\left(\rho_{ir}b_{o}^2 f_{3}^5 +\frac{1}{f_3}\rho_{inr}\right)
\end{equation}
If we make the assumption as before that $\dot{\phi}$ = 0 = $\dot{b}$ this equation will give the RFW relation for the Hubble parameter but now with matter from three separate branes included.  However,the situation with no matter included on the 5-brane
in the bulk does not reduce to the case with only two
branes since $\alpha$ and therefore the function f(y) are
modified from the case where only two branes were present.  Therefore, the Hubble law (more specifically the Newton constant,
$G_{N}$) is affected by the presence of the additional
five brane even if it is empty.   

The other field equations can be solved in a manner similar to that described above for the $G_{tt}$ equation.  Namely,
we first modify the equations with the inclusion of terms involving the y-derivative of $\alpha$, look at the 
equations separately in the regions, 0$\leq$ y $<$ Y and Y $<$ y $\leq$ $\pi$$\rho$,  and then match them across the 
five brane.  The result for the $\phi$ equation in the bulk is
\begin{equation}
\Delta V'' +\frac{3\alpha}{f}\left(\Delta n' +3\Delta a' - \Delta V' \right)
=b_{o}^2f^3A_4 + \frac{1}{M_{5}^3 f_3}\rho_{3nr}\delta\left(y-Y\right)
\end{equation}  
which leads to
\begin{equation}
\int_{0}^{\pi\rho}dy'b_{o}^2 f^3\left(A_4 - 3A_3\right) = -\sum_{i=1}^3\frac{1}{M_{5}^3 f_i}\rho_{inr}.
\end{equation}
The $G_{kk}$ equation gives
\begin{equation}
\int_{0}^{\pi\rho}dy'b_{o}^2 f^3\left(A_1  + A_2\right) = \sum_{i=1}^3\frac{1}{6M_{5}^3 f_i}\rho_{inr}.
\end{equation}
The $G_{yy}$ equation remains unchanged from Eq.(25) since there are no matter sources present but where now the gauge invariant combinations $\Delta a$, $\Delta n$, and $\Delta V$ include the new definitions of $\alpha$(y)
and f(y).  We can now proceed to check the consistency using the same steps that led to Eq.(41).  The new relation is
\begin{equation}
\left(\frac{1}{f_1}\rho_{1nr}+\frac{1}{f_2}\rho_{2nr}+\frac{1}{f_3}\rho_{3nr}\right) = 0
\end{equation}
Once again we see that introducing non-relativistic matter into the system results in an inconsistency if there is no fine tuning
of the matter on the different branes.   

\section{y-Dependence and Non-Relativistic Matter}
Assuming the static case for $\phi$ and b we have shown that the system does not admit consistent solutions in the presence of non-relativistic matter.  We would now like to relax this assumption and then examine the system.  We consider here the case of no 5-branes present.  To see what constraints are put on our assumptions, let us first look at two separate ways of evaluating the y-derivative of H and compare this with the $G_{ty}$ equation.  The definition of H is
\begin{equation}
H \equiv \frac{\dot{a}}{a}.
\end{equation}
Therefore 
\begin{equation}
H' = \frac{\dot{a}'}{a}- \frac{a'}{a}\frac{\dot{a}}{a}
\end{equation}
and using the explicit form of $a(y,t)$ in Eq.(15) we find
\begin{equation}
\frac{\dot{a}'}{a} = \left(\frac{\alpha}{2f}+\delta a'\right)H + H'.
\end{equation}
Alternately
\begin{equation}
\frac{a'}{a} = \frac{\alpha}{2f} + \delta a'
\end{equation}
which gives
\begin{equation}
\frac{\dot{a}'}{a} = \left(\frac{\alpha}{2f}+\delta a'\right)H + \delta\dot{a}'.
\end{equation}
Comparing Eq.(67) with Eq.(69) and using the fact that in the static case $\Delta\dot{a}'$ = $\delta\dot{a}'$ we find
\begin{equation}
\Delta\dot{a}' = H'.
\end{equation}
In the static case the $G_{ty}$ equation is given by
\begin{equation}
\Delta\dot{a}' = \left(\Delta n' - \Delta a'\right)H.
\end{equation}
The right hand side is seen to be of order $\rho^{3/2}$ which shows from Eq.(70) that $H'$ is also of order $\rho^{3/2}$.  Thus the static case $\emph{requires}$ H to be independent of y to first order in $\rho$.

Let us next examine the situation when we let $\phi$ and b depend on time.  The Einstein equations $G_{tt}$, $G_{kk}$ + $G_{tt}$, $G_{yy}$, and $G_{yy}$ plus the $\phi$ equation of motion (Eqs.(35),(37),(39),(40)) give
\begin{eqnarray}
\int_{f_1}^{f_2}df'f'^3 A_1 &=& \frac{\lambda}{6}\sum_{i=1}^{2}\left(b_{o}^2 \rho_{ir}f_{i}^5 + \frac{\rho_{inr}}{f_i}\right)\\
\int_{f_1}^{f_2}df'f'^3 (A_1+A_2) &=& \frac{\lambda}{6}\sum_{i=1}^{2}\frac{\rho_{inr}}{f_i}\\
\int_{f_1}^{f_2}df'f'^3 (3A_3-A_4) &=& \lambda\sum_{i=1}^{2}\frac{\rho_{inr}}{f_i}\\
\int_{f_1}^{f}df'f'^3 (A_1 +A_2+3A_3-A_4) &=& f^4 A_3(y) + \frac{7\lambda}{6}\frac{\rho_{1nr}}{f_1}
\end{eqnarray}
In particular evaluating Eq.(75) at y = $y_1$ gives
\begin{equation}
A_3(y_1) = -\frac{7\lambda\rho_{1nr}}{6f_{1}^{5}}.
\end{equation}
 However, considering the combination Eq.(73) + Eq.(74) - Eq.(75), where Eq.(75) is evaluated at y = $y_2$, yields
\begin{equation}
A_3(y_2) = \frac{7\lambda\rho_{2nr}}{6f_2^5}
\end{equation}
and therefore we see that $A_3$ has non-trivial y-dependence.  Since $A_3$ only depends on $H$ and $\dot{\phi}$ this implies that $\dot{\phi}$ depends on y.  With the static condition relaxed we can now see that the Einstein equations contain integrals over functions whose y-dependence is not known.  Therefore relations such as Eq.(41) are no longer valid.  Eq.(41) now becomes
\begin{equation}
\int_{f_1}^{f_2}df'\left(9H\frac{\dot{b}}{b} + 3\frac{\ddot{b}}{b} + \dot{\phi}^2 + \ddot{\phi} + \frac{\dot{b}}{b}\dot{\phi} + 3H\dot{\phi}\right) = -\frac{\lambda}{2}\left(\frac{1}{f_1}\rho_{1nr} + \frac{1}{f_2}\rho_{2nr}\right)
\end{equation}
and we see that non-relativistic matter is now related to an integral over an unknown function of y.  Without knowing the exact form of the integrand it is difficult to determine if there is any constraint on the non-relativistic matter with the static constraint relaxed.    

\section{Randall-Sundrum Model}
The Randall Sundrum Model (RS1) \cite{rs} is a phenomenological five dimensional model with the fifth dimension bounded by two 3-branes, and thus it resembles the 5D reduction of Horava-Witten M-theory.  It is of interest then to compare the two to see what differences exist.  A problem arose in the earlier versions of RS1 when it was realised that the correct Hubble era cosmology could not be reproduced, and it was then suggested that one could obtain the RWF cosmology if one were to stabilize the fifth dimension \cite{csaki}.  A technique for this was proposed by Goldberger and Wise \cite{gw} by adding a scalar field in the bulk with appropriate bulk and brane potentials, and this idea was further analyzed by Cline and Firouzjahi \cite{cline2}.  However, neither Refs. \cite{gw} or \cite{cline2} included the dynamics of the gravitational field in the bulk.  

In general it is very difficult to find vacuum solutions for the Einstein-scalar field equations with arbitrary brane and bulk potentials.  Further, we have seen in Sec.3 how important it is to have a rigorous vacuum solution since it is the careful imposition of boundary conditions on both branes that produces tension in the system.  However, a special class of solutions for the vacuum metric obeying all boundary conditions on the branes was constructed by deWolfe et. al \cite{dewolfe} where the brane and bulk potentials are related to a single function of the scalar field.  Subsequently, in Ref.\cite{cline}, matter on the branes was added and treated perturbatively with respect to a vacuum solution.  However, an explicit vacuum solution for their choice of potentials was not obtained, and the difficulty in obtaining a solution is discussed in the Appendix.  Here we will discuss the case of \cite{dewolfe} with matter treated as a perturbative addition on the branes , since this case treats the vacuum solution rigorously.  

In order to make contact with the notation found in \cite{dewolfe} we let $V(\phi)$ be the bulk potential and $\lambda_{i}(\phi)$, i=1,2 be the potentials on the two branes at y=$y_1$ and y=$y_2$ respectively.  The metric is given by
\begin{equation}
ds^2 = e^{2N(t,y)}dt^2 - e^{2A(t,y)}\sum_{i}dx_{i}^{2} - b(t,y)^{2}dy^2
\end{equation}
with the perturbative expansions
\begin{eqnarray}
N(t,y) &=& A_o(y) + \delta N(t,y)\\
A(t,y) &=& A_o(y) + \delta A(t,y)\\
b(t,y) &=& 1 + \delta b(t,y)\\
\phi(t,y) &=& \phi_{o}(y) + \delta\phi(t,y).
\end{eqnarray}
We will be working in the static case where $\dot{\phi}$ = 0 = $\dot{b}$.

The vacuum equations in the bulk are
\begin{eqnarray}
\phi_{o}'' + 4A_{o}'\phi_{o}' = V'(\phi_{o})\\
A_{o}'' = -\frac{2}{3}\phi_{o}'^{2}\\
A_{o}'^{2} = -\frac{1}{3}V(\phi_{o}) + \frac{1}{6}\phi_{o}'^2
\end{eqnarray}
where primes denote $\frac{\partial}{\partial y}$ except on $V(\phi)$ where it represents $\frac{\partial}{\partial\phi}$.  The boundary conditions are given by 
\begin{eqnarray}
A_{o}'\bigg|_{y=y_{i}} = (-1)^{i}\frac{b(y)}{3}\lambda_{i}(\phi_{o})\bigg|_{y=y_{i}}\\
\phi_{o}'\bigg|_{y=y_{i}} = (-1)^{i+1}\frac{b(y)}{2}\frac{\partial\lambda_{i}(\phi_{o})}{\partial\phi_{o}}\bigg|_{y=y_{i}}
\end{eqnarray}
where i = 1,2 corresponds to the brane locations at y = $y_1$, $y_2$.

It was shown in \cite{dewolfe} that if $V(\phi_{o})$ has the form
\begin{equation}
V(\phi_{o}) = \frac{1}{8}\left(\frac{\partial W(\phi_{o})}{\partial\phi_{o}}\right)^2 - \frac{1}{3}W(\phi_{o})^2
\end{equation}
for some $W(\phi_{o})$, then a solution of the bulk vacuum equations will also satisfy
\begin{equation}
\phi_{o}' = \frac{1}{2}\frac{\partial W(\phi_{o})}{\partial\phi_{o}}\,\,\,,\,\,\,A_{o}' = -\frac{1}{3}W(\phi_{o})
\end{equation}
as long as the boundary conditions
\begin{equation}
W(\phi_{o})\bigg|_{y=y_{i}} = (-1)^{i+1}\lambda_{i}(\phi_{o})\bigg|_{y=y_{i}}\,\,\,,\,\,\,\frac{\partial W(\phi_{o})}{\partial\phi_{o}}\bigg|_{y=y_{i}} = (-1)^{i+1}\frac{\partial \lambda_{i}(\phi_{o})}{\partial\phi_{o}}\bigg|_{y=y_{i}}
\end{equation}
are also satisfied.  Once $W(\phi_{o})$, $\lambda_{1}$, and $\lambda_{2}$ are chosen one is then left with a set of first order differential equations that can easily be solved. 

In the example constructed in \cite{dewolfe} $W(\phi_{o})$, $\lambda_{1}$, and $\lambda_{2}$ were chosen to be
\begin{eqnarray}
W(\phi_{o}) &=& \frac{3}{L} - b\phi_{o}^2\\\nonumber
\lambda_{1}(\phi_{o}) &=& W(\phi_{o}(y_1))+W'(\phi_{o}(y_1))(\phi_{o} - \phi_{o}(y_1)) \\ &&+ \gamma_{1}(\phi_{o} - \phi_{o}(y_1))^2\\\nonumber
\lambda_{2}(\phi_{o}) &=& -W(\phi_{o}(y_2))-W'(\phi_{o}(y_2))(\phi_{o} - \phi_{o}(y_2)) \\ &&+ \gamma_{2}(\phi_{o} - \phi_{o}(y_2))^2
\end{eqnarray}
which gives
\begin{eqnarray}
\phi_{o}(y) &=& \phi_{1}e^{-\beta y}\\
A_{o}(y) &=& a_o - \frac{y}{L} - \frac{1}{6}\phi_{1}^{2}e^{-2\beta y}
\end{eqnarray}
where $a_o$, $L$, $\gamma_1$, $\gamma_2$, $\phi_1$, and $\phi_2$ are all arbitrary parameters.  The relationship between $\lambda_i(\phi_o)$ and $W(\phi_o)$ fine tunes the net cosmological constant on the branes to zero.  While there is no a priori motivation for the choice (other than the fact it fine tunes the cosmological constant to zero), it leads to simple analytic forms, Eqs.(95),(96), which can be treated easily.

Next we define the quantities
\begin{eqnarray}
\Delta A' &=& \delta A' + \frac{2}{3}\phi_{o}'\delta\phi -A_{o}'\delta b\\
\Delta N' &=& \delta N' + \frac{2}{3}\phi_{o}'\delta\phi -A_{o}'\delta b\\
\Delta V' &=& \phi_{o}'\delta\phi' - \phi_{o}''\delta\phi -\phi_{o}'^2\delta b.
\end{eqnarray}
These quantities are invariant under a first order coordinate change in the y coordinate: $\bar{y}$ = y + $\delta$(y).

To first order in the matter the Einstein equations $G_{tt}$, $G_{kk}$, $G_{yy}$, and the field equation for $\phi$ in the bulk become
\begin{eqnarray}
\Delta A'' + 4A_{o}'\Delta A' = e^{-2A_{o}(y)}H^2 \equiv A_1\\
2\Delta A'' + 8A_{o}'\Delta A' + \Delta N'' + 4A_{o}'\Delta N' = e^{-2A_{o}(y)}\left(3H^2 + 2\dot{H}\right) \equiv A_2\\
A_{o}'\left(3\Delta A' + \Delta N'\right) -\frac{2}{3}\Delta V' = e^{-2A_{o}(y)}\left(2H^2 + \dot{H}\right) \equiv A_3\\
\Delta V'' + 4A_{o}'\Delta V' + \phi_{o}'^{2}\left(3\Delta A' + \Delta N'\right) = 0.
\end{eqnarray}
The boundary conditions are
\begin{eqnarray}
\Delta A'\bigg|_{y=y_{i}} = \left(-1\right)^{i}\frac{\rho_{i}}{3}\\
\left(3\Delta A' + \Delta N'\right)\bigg|_{y=y_{i}} = \left(-1\right)^{i}\frac{\rho_{inr}}{3}\\
\Delta V'\bigg|_{y=y_{i}}= \delta\phi_{i}\left(\left(-1\right)^{i+1}\gamma_{i}\phi_{o}'-\phi_{o}''\right)\bigg|_{y=y_{i}}
\end{eqnarray}
where $\gamma_{i}$ are the free parameters in the model of Eqs.(93),(94) i.e.
\begin{equation}
\gamma_i = {1\over 2}\frac{\partial^{2}\lambda_i}{\partial\phi^2}.
\end{equation}

Immediately we see a difference between HW theory and the RS phenomenology manifesting itself in the boundary conditions for $\Delta V'$.  In HW only the invariant quantity $\Delta V'$ appears in the boundary conditions whereas an additional $\delta\phi$ term appears in the RS model.  One can easily check that in HW there is no free parameter analogous to $\gamma_{i}$, and the M-theory choice of potential makes the term in parenthesis on the right-hand side of Eq.(106) zero.  This distinction will allow one to avoid the constraint on matter found in the HW theory.

We will now solve the system in the presence of arbitrary matter.  The $G_{tt}$ equation can be integrated to obtain
\begin{equation}
\Delta A' = e^{-4A_{o}(y)}\int_{y_1}^{y}dy'e^{2A_o(y')}H^2 + e^{-4A_{o}(y)}c_1
\end{equation}
and using the boundary conditions we get 
\begin{eqnarray}
c_1 &=& -\frac{\rho_1}{3}e^{4A_{o}(y_1)}\\
\int_{y_1}^{y_2}dy'e^{2A_{o}(y')}H^2 &=& \frac{\rho_1}{3}e^{4A_{o}(y_1)} + \frac{\rho_2}{3}e^{4A_{o}(y_2)}
\end{eqnarray}
If H is independent of y one can see that the usual Friedman relation is recovered.  Eq.(108) now becomes
\begin{equation}
\Delta A' = e^{-4A_{o}(y)}\left(\int_{y_1}^{y}dy'e^{2A_o(y')}H^2 - \frac{\rho_1}{3}e^{4A_{o}(y_1)}\right)
\end{equation}
In a similar manner we can solve the combination $G_{tt}$ + $G_{ii}$ to obtain
\begin{eqnarray}
\int_{y_1}^{y_2}dy'e^{2A_{o}(y')}(4H^2 + 2\dot{H}) = \frac{\rho_{1nr}}{3}e^{4A_{o}(y_1)} + \frac{\rho_{2nr}}{3}e^{4A_{o}(y_2)}\\
3\Delta A' + \Delta N' = e^{-4A_{o}(y)}\left(\int_{y_1}^{y}dy'e^{2A_{o}(y')}(4H^2 + 2\dot{H})-\frac{\rho_{1nr}}{3}e^{4A_{o}(y_1)}\right).
\end{eqnarray}
The remaining equation is $G_{yy}$ which can be solved for $\Delta V$ or equivalently for $\delta\phi(y)$ in terms of $\delta b(y)$
\begin{eqnarray}\nonumber
\delta\phi(y) = \phi_{o}'\int_{y_1}^{y}dy'\delta b(y') + \phi_{o}'\int_{y_1}^{y}dy'\frac{A_{o}'}{\phi_{o}'^2}\left(3\Delta A' + \Delta N'\right)\\ - \phi_{o}'\int_{y_1}^{y}dy'\frac{e^{-2A_{o}(y')}}{\phi_{o}'^2}\left(2H^2 + \dot{H}\right) + \delta\phi(y_1).
\end{eqnarray}
The boundary conditions on $\phi(y)$ are\footnote{We note the limit $\gamma_i$ $\to$ $\infty$ is the stiff potential limit of \cite{cline}}
\begin{eqnarray}
\delta\phi(y_{i}) = \frac{\frac{(-1)^i}{2}A_{o}'(y_{i})\rho_{inr} - e^{-2A_{o}(y_{i})}\left(2H^2+\dot{H}\right)}{(-1)^{i+1}\gamma_{i}\phi_{o}'(y_{i}) - \phi_{o}''(y_{i})}
\end{eqnarray}
One can then substitute the known expressions for $A_{o}'$, $\phi_{o}'$ (for the specific model of \cite{dewolfe} these are given by Eqs.(95),(96)), and 2$H^2$ + $\dot{H}$ from Eq.(112) into Eq.(115).  This then determines $\delta\phi(y_i)$ in terms of $\rho_{inr}$.  However, there remains one additional relation involving $\delta\phi(y_i)$ for we may let y = $y_2$ in Eq.(114).  Eliminating $\delta\phi(y_1)$ and $\delta\phi(y_2)$ using Eq.(115), then Eq.(114) at y = $y_2$ becomes a constraint on $\delta b(y)$:
\begin{eqnarray}\nonumber
\int_{y_1}^{y_2}dy'\delta b(y') &=& \frac{\delta\phi(y_{2})-\delta\phi(y_{1})}{\phi_{o}'} + \int_{y_1}^{y_2}dy'\frac{e^{-2A_{o}(y')}}{\phi_{o}'^2}\left(2H^2 + \dot{H}\right)  \\ && - \int_{y_1}^{y_2}dy'\frac{A_{o}'}{\phi_{o}'^2}\left(3\Delta A' + \Delta N'\right).
\end{eqnarray}
Note that the right hand side of Eq.(116) depends only on the vacuum metric and $\rho_{inr}$ which can be seen upon substitution of Eqs.(112),(113),(115)
\begin{eqnarray}\nonumber
&&\int_{y_1}^{y_2}dy'\delta b(y') = \\\nonumber&&\frac{\frac{A_{o}'(y_2)}{2}\rho_{2nr} - e^{-2A_{o}(y_2)}F^{-1}\sum_{i}\frac{\rho_{inr}}{6}e^{4A_{o}(y_i)}}{-\gamma_{2}\phi_{o}'^{2}(y_2) - \phi_{o}''(y_2)\phi_{o}'(y_2)}\\\nonumber&+& \frac{\frac{A_{o}'(y_1)}{2}\rho_{1nr} + e^{-2A_{o}(y_1)}F^{-1}\sum_{i}\frac{\rho_{inr}}{6}e^{4A_{o}(y_i)}}{\gamma_{1}\phi_{o}'(y_1)\phi_{o}'(y_2) - \phi_{o}''(y_1)\phi_{o}'(y_2)}\\&+&\int_{y_1}^{y_2}dy'\frac{e^{-2A_{o}(y')}}{\phi_{o}'^2(y')}F^{-1}\sum_{i}\frac{\rho_{inr}}{6}e^{4A_{o}(y_i)}\\\nonumber&+&\int_{y_1}^{y_2}dy'\frac{A_{o}'(y')}{\phi_{o}'^2(y')}e^{-4A_o(y')}\frac{\rho_{1nr}}{3}e^{4A_{o}(y_1)}\\\nonumber
&-&\int_{y_1}^{y_2}dy'\frac{A_{o}'(y')}{\phi_{o}'^2(y')}e^{-4A_o(y')}\int_{y_1}^{y'}dy''e^{2A_{o}(y'')}F^{-1}\sum_{i}\frac{\rho_{inr}}{3}e^{4A_{o}(y_i)}
\end{eqnarray}
where
\begin{equation}
F = \int_{y_1}^{y_2}dy'e^{2A_o(y')}.
\end{equation}
Therefore a consistent solution for arbitrary non-relativistic matter is obtained. (It should be noted that although these equations hold in general, explicit values for $\delta\phi(y_i)$ can only be calculated if $A_{o}'$ and $\phi_{o}'$ can be determined, which will be dependent on the choice of the bulk potential.)    

We see now the meaning of the result that the $\delta\phi$ boundary condition depends on both the coordinate invariant combination $\Delta V'$ \emph{and} on $\delta\phi_i$ (rather than just on $\Delta V'$ as in HW).  Instead of putting a constraint on $\rho_{inr}$, this determines the integral of $\delta b(y)$ in Eq.(116) which is just the change of distance between the branes due to the presence of non-relativistic matter.  This is possible with the phenomenological potentials of the RS model, but not in HW where the theory determines the potentials to automatically cancel the cosmological constant.  Since $\rho_{inr}$ decreases as 1/$a(t)^3$, the brane separation is actually time dependent at higher order in RS so the distance between the branes cannot be fixed, and so the RS model is also non-static.  Here, however, the time variation of the invariant distance between the branes sets in at $\cal{O}$($\rho_{inr}^{3/2}$) since $\dot{\rho} \sim H\rho$ = $\cal{O}$($\rho^{3/2}$) while in HW the time variation sets in at $\cal{O}$($\rho$).

The above illustrates the difference between the phenomenology of RS and the theory of HW.  In the RS model one is free to add arbitrary bulk and brane potentials (characterised here by $\gamma_i$) for the scalar field while in HW the couplings of the volume modulus $V$ are determined by the theory and one is not free to include ad hoc potentials.  In our analysis of HW theory we have used all potentials that arise perturbatively.  There are non-perturbative potentials in HW that have not been included that might relax the tension that non-relativistic matter produces.  If this is the case, one would also have to modify the vacuum solution to take into account the additional interactions.

\section{Conclusions}
We have studied here the Hubble era cosmology in Horava-Witten theory.  After compactification of the 11D space on a Calabi-Yau threefold, the system reduces to a 5D theory, the fifth dimension, y, bounded by two 3-branes with gravity and a scalar field (representing the volume modulus) in the bulk, and gauge and chiral matter on the 3-branes.  The field equations were solved in the bulk and the boundary conditions on both 3-branes were imposed.  We have shown that for the static solution, the standard RWF cosmology arises for relativistic matter on the branes, but the field equations cannot allow non-relativistic matter.  This result arises from the constraint of satisfying the boundary conditions on $\emph{both}$ 3-branes.  The same result maintains if one adds 5-branes in the bulk (the most general form of Horava-Witten M theory).  We have included all of the potentials that arise perturbatively in HW theory.  However, there are non-perturbative potentials in the theory that have not been included and may allow the introduction of non-perturbative matter.  Once these potentials are included, the vacuum structure of the theory will be altered and one will need to find a new set of vacuum solutions.  We also have not addressed the issue of moduli stabilization that has been studied recently in the context of flux compactifications in \cite{buchbinder,becker}.   

We have also demonstrated the difference between the HW theory and the RS phenomenology.  For the special class of potentials studied in \cite{dewolfe} the RS model can accommodate arbitrary matter on the branes and thereby reproduce the RWF cosmology.  This occurs due to the existence of free parameters in the brane potentials which can relax the constraint on non-relativistic matter found for HW theory by allowing one to solve for the change in brane separation due to the presence of matter rather than putting a constraint on the matter itself.  In HW theory one is not allowed to introduce such free parameters as the form of the brane potentials are fully determined by the consistency conditions of the theory.    

In the Appendix we have analysed the RS model discussed in \cite{cline} which includes a scalar potential that does not fall into the class of potentials studied in \cite{dewolfe}.  The vacuum solutions in this case are given by a series expansion and we have shown that one cannot generate the desired solution to the hierarchy problem if one truncates the series after the first term as was done in \cite{cline} (or if one includes the second term).  The difficulty is that the vacuum solution must obey boundary conditions on both branes which precludes the hierarchy from developing.  It is an interesting question of whether this difficulty is due to the truncation or whether formation of a hierarchy is sensitive to the choice of brane and bulk potentials.  

\section{Acknowledgments}
This work is supported in part by a National Science Foundation Grant
PHY-0101015 and in part by  Natural Sciences and Engineering Research Council 
of Canada. 

\appendix
\section{Appendix}

\renewcommand{\theequation}{\Alph{section}.\arabic{equation}}
\setcounter{equation}{0}

In section 6 we discussed the Randall-Sundrum model and showed that for a general class of bulk and brane potentials one can obtain vacuum solutions and then find consistent solutions to the Einstein and scalar field equations for arbitrary matter introduced perturbatively on the branes.  In this appendix we will analyse the situation for a scalar field obeying the potentials chosen in \cite{cline},i.e. in the bulk
\begin{equation}
V(\Phi) = \frac{1}{2}m\Phi^2,
\end{equation}  
and on the boundaries $V_{i}(\Phi)=m_{i}(\Phi-v_{i})^2$.  Here $m_{i}$ are the analogs of $\gamma_{i}$ that arose in Sec 6.  These potentials cannot be put into the form specified by Eq.(90) and therefore one will have to solve the second order differential equations to find the vacuum solutions.  We will find that it is not easy to see how this choice of potentials leads to the desired hierarchy.  

Throughout this section we follow the notation of \cite{cline}.  The metric is
\begin{equation}
ds^2 = e^{-2N(t,y)}dt^2 - e^{-2A(t,y)}\sum_{i}dx_{i}^{2} - b(t,y)^{2}dy^2
\end{equation}
and the perturbative expansions are given by
\begin{eqnarray}
N(t,y) &=& A_o(y) + \delta N(t,y)\\
A(t,y) &=& A_o(y) + \delta A(t,y)\\
b(t,y) &=& b_o + \delta b(t,y)\\
\Phi(t,y) &=& \Phi_{o}(y) + \delta\Phi(t,y).
\end{eqnarray}
The Einstein equations and the scalar field equation at vacuum order are
\begin{eqnarray}
A_{o}'^2 &=& \frac{\kappa^2}{12}\left(\Phi_{o}'^2 - m^{2}b_{o}^2\Phi_{o}^2\right) +k^{2}b_{o}^2\\
A_{o}'' &=& \kappa^2\frac{1}{3}\Phi_{o}'^2\\
\Phi_{o}'' &=& 4A_{o}'\Phi_{o}' +m^{2}b_{o}^2\Phi_{o}
\end{eqnarray}
where $\kappa^2$ is given by $\kappa^2$ = 1/$M^3$ where $M$ is the 5D Planck scale and $\kappa^2$ and $k^2$ are related to the bulk cosmological constant $\Lambda$ by $\Lambda$ = -6$k^2$/$\kappa^2$.  Only two of these equations are independent since the third equation can be generated by taking the y-derivative of the first and then inserting the second.  Therefore a solution of any two of these equations will necessarily satisfy the third.  We look for solutions of the form
\begin{eqnarray}
A_o &=& a_o + \beta y + \sum_{n=1}^{\infty}a_{n}e^{-2n\alpha y}\\
\Phi_{o} &=& \sum_{n=1}^{\infty}c_{n}e^{-(2n-1)\alpha y}
\end{eqnarray}
where $\alpha$ = $\epsilon k b_{o}$.  

Ref.\cite{cline} assumed that truncating the series at n=1 represents a good approximation.  However, it is easy to see that when this truncation is inserted into Eqs.(A.7-A.9) one generates the higher terms of Eqs.(A.10,A.11).  We would like to examine the question of whether the higher terms can be ignored thereby giving the results found in \cite{cline}.  While the equations are non-linear, it is still possible to obtain recursion relations.  Inserting Eqs.(A.10) and (A.11) into the vacuum equation (A.8) generates relations between $a_{n}$ and $c_{n}$, the first few of which are
\begin{eqnarray}
4a_1 &=& \frac{\kappa^2}{3}c_{1}^2\\
16a_2 &=& \frac{\kappa^2}{3}6c_{1}c_{2}\\
36a_3 &=& \frac{\kappa^2}{3}\left(10c_{1}c_{3}+9c_{2}^2\right).
\end{eqnarray}
One can also see that $\beta = kb_{o}$ from Eq.(A.7).
Inserting Eqs.(A.10) and (A.11) into Eq.(A.9) we find
\begin{eqnarray}
\sum_{n=1}^{\infty}c_{n}\left(\alpha^2\left(2n-1\right)^2 + 4\beta\alpha\left(2n-1\right)-m^{2}b_{o}^2\right)e^{-(2n-1)y} \\= 8\alpha^{2}\sum_{n,m=1}^{\infty}n(2m-1)a_{n}c_{m}e^{-(2n+2m-1)y}.
\end{eqnarray} 
For n=1 this gives an equation for the coefficients of $e^{-\alpha y}$,
\begin{eqnarray}
\alpha^2 + 4\alpha\beta -m^{2}b_{o}^2 = 0
\end{eqnarray}
or
\begin{eqnarray}
\epsilon = -2 + \sqrt{4 + \frac{m^2}{k^2}}
\end{eqnarray}
where we have taken the positive root so that $\epsilon > 0$.(This is identical to the result found in \cite{cline}.)  For n = 2, after using the result of Eq.(A.12), we find the following relation between $c_2$ and $c_1$
\begin{equation}
c_2 = \frac{2\alpha^2\kappa^2}{3(9\alpha^2 + 12\beta\alpha -m^{2}b_{o}^2)}c_{1}^3.
\end{equation}
Using Eq.(A.17) reduces this to 
\begin{equation}
c_2 = \frac{1}{12}\frac{\alpha\kappa^2}{\alpha + \beta}c_{1}^3
\end{equation}
which is an example of the general result 
\begin{equation}
c_n \sim \kappa^{2n-2}c_{1}^{2n-1}
\end{equation}
and subsequently
\begin{equation}
a_n \sim \kappa^{2n}c_{1}^{2n}.
\end{equation}
Thus all the coefficients are determined by the constant of integration $c_1$.

Solutions to the bulk equations must also satisfy the boundary conditions
\begin{eqnarray}
A_{o}'\bigg|_{y=y_{i}} = (-1)^{i+1}\frac{\kappa^2}{6}b_{o}V_{i}(\Phi)\bigg|_{y=y_{i}}\\
\Phi_{o}'\bigg|_{y=y_{i}}= (-1)^{i+1}\frac{b_o}{2}V_{i}'(\Phi)\bigg|_{y=y_{i}}
\end{eqnarray}
where i=1,2 refers to the boundary at y=0, and y=1 respectively and $V_i$ is specified below Eq.(A.1).  We will first try to satisfy these boundary conditions by keeping only the first term in the series for $\Phi_o$ and $A_o$ and then use the relation Eq.(A.19) to determine if this is a reasonable approximation.  Using
\begin{eqnarray}
\Phi_{o}' = -\alpha c_{1}e^{-\alpha y}\\
A_{o}' = \beta -\frac{\kappa^2}{6}\alpha c_{1}^2 e^{-2\alpha y}
\end{eqnarray}  
in the $\Phi_o$ boundary conditions we find equations for $c_1$ and $e^{-\alpha}$ in terms of $v_1$ and $v_2$
\begin{eqnarray}
c_1 = \frac{m_1 v_1}{m_1 + k\epsilon}\\
e^{-\alpha} = \frac{m_2 v_2}{m_2 - k\epsilon}\frac{m_1 + k\epsilon}{m_1 v_1}.
\end{eqnarray}
Note that if $m_{1,2}$ $\gg$ $k\epsilon$ (the ``stiff potential'' limit) then
\begin{eqnarray}
e^{-\epsilon k b_o} \simeq \frac{v_2}{v_1}
\end{eqnarray}
which was obtained in \cite{cline} and used there to create a large hierarchy without the need for fine-tuning of parameters.  Thus writing Eq.(A.29) as
\begin{equation}
e^{-kb_o} \simeq \left(\frac{v_2}{v_1}\right)^{1/\epsilon}
\end{equation}
one see that for e.g. $\epsilon$ = $1/30$ one needs only assume $v_2/v_1$ = 0.3 to obtain
\begin{equation}
e^{-kb_o} \cong 2\times10^{-16}
\end{equation}
However, $v_1$ and $v_2$ are not totally free parameters and using the boundary conditions for $A_o$ we find relations for $v_1$ and $v_2$
\begin{eqnarray}
v_{1}^2 = \frac{6}{\kappa^2}\frac{m_1 + \epsilon k}{\epsilon m_1}\\
v_{2}^2 = \frac{6}{\kappa^2}\frac{m_2 - \epsilon k}{\epsilon m_2}
\end{eqnarray}
When these relations are inserted into Eq.(A.28) we find that
\begin{eqnarray}
e^{-\alpha} = \left(\frac{m_2}{m_2 - k\epsilon}\frac{m_1 + k\epsilon}{m_1}\right)^{1/2}.
\end{eqnarray}
This implies that $e^{-\alpha}$ $\geq$ 1 and therefore does not give a solution to the hierarchy problem.  It should be noted that this result holds for any $m_i$.

One can also see that with $\epsilon$ $\ll$ 1 the n=2 term in the series is not small compared to the n=1 term.  From Eq.(A.19) the ratio $c_2$/$c_1$ becomes
\begin{eqnarray}
\frac{c_2}{c_1} \simeq \frac{\epsilon}{12}(\kappa c_1)^2
\end{eqnarray}
when $\epsilon$ $\ll$ 1.  Using Eq.(A.27) for $c_{1}^2$ and Eq.(A.32) for $v_{1}^2$ we find 
\begin{eqnarray}
\frac{c_2}{c_1} \simeq \frac{m_1}{2(m_1 + k\epsilon)}
\end{eqnarray}
which is of $\mathcal{O}$(1),  and in the stiff potential limit $c_2$/$c_1$ $\simeq$ 1/2.  Therefore truncating the series to the first term is not a valid approximation, as the small parameter $\epsilon$ in Eq.(A.35) cancels out in Eq.(A.36).  It is the imposition of the $A_{o}'$ boundary condition that makes $v_{1,2}$ $\sim$ $1/\epsilon^{1/2}$.   

We next examine the effect of retaining only the first two terms in the series expansions for $\Phi_o$ and $A_o$:
\begin{eqnarray}
A_{o} = a_o + kb_o + a_{1}e^{-2\alpha y} + a_{2}e^{-4\alpha y}\\
\Phi_{o} = c_{1}e^{-\alpha y} + c_{2}e^{-3\alpha y}.
\end{eqnarray}
As was previously noted, all coefficients can be found in terms of $c_1$.  The $\Phi_{o}$ boundary condition at y=0 becomes
\begin{equation}
\tilde{c_1}^3 + \frac{1 + \delta_1}{1 + 3\delta_1}\tilde{c_1} - \frac{1}{1 + 3\delta_1}\tilde{v_1} = 0
\end{equation}
where we have introduced the notation
\begin{eqnarray}
\tilde{c}_1 &\equiv& \left(\frac{\epsilon\kappa^2}{12(1+\epsilon)}\right)^{1/2}c_1\\
\tilde{v}_1 &\equiv& \left(\frac{\epsilon\kappa^2}{12(1+\epsilon)}\right)^{1/2}v_1\\
\delta_{i}  &\equiv& \frac{k\epsilon}{m_i}.
\end{eqnarray}
Inserting Eqs.(A.37) and (A.38) into the y=0 boundary condition for $A_o'$ gives
\begin{eqnarray}
\frac{\delta_1}{2(1+\epsilon)} - \delta_{1}\tilde{c}_1\left(\tilde{c}_1+3\tilde{c}_1^3\right) = \left(\tilde{c}_1+\tilde{c}_1^3 -\tilde{v}_1\right)^2.
\end{eqnarray}
Upon substituting for $\tilde{c}_1^3$ from Eq.(A.39), Eq.(A.43) becomes
\begin{equation}
2(1+\delta_1)\tilde{c}_1^2 -3(1-\delta_1)\tilde{c}_1\tilde{v}_1+\frac{(1+3\delta_1)^2}{2(1+\epsilon)}-9\delta_{1}\tilde{v}_1^2=0.
\end{equation}
This is easily solved for $\tilde{c}_1$ in terms of $\tilde{v}_1$:
\begin{eqnarray}
\tilde{c}_1 = \frac{3\tilde{v}_1}{4(1+\delta_1)}\left(1-\delta_1 \pm \left[(3\delta_1 + 1)^2 -\frac{4(1+\delta_1)(1+3\delta_1)^2}{9\tilde{v}_1^2(1+\epsilon)}\right]^{1/2}\right).
\end{eqnarray}
In the stiff potential limit, $\delta_1$ $\to$ 0, this reduces to 
\begin{eqnarray}
\tilde{c}_1= \frac{3\tilde{v}_1}{4}\left(1 \pm \left[1-\frac{4}{9\tilde{v}_1^2(1+\epsilon)}\right]^{1/2}\right).
\end{eqnarray}
Taking the positive root and putting this into Eq.(A.39) leads to an equation for $\tilde{v}_1$ that can be solved to give
\begin{equation}
\tilde{v}_1 \cong .667.
\end{equation}
No real solution is found if one takes the negative root in Eq.(A.46).  

From the y=1 boundary conditions we obtain the equations
\begin{equation}
e^{-3\alpha}\tilde{c}_1^3 + \frac{1 - \delta_2}{1 - 3\delta_2}e^{-\alpha}\tilde{c}_1 - \frac{1}{1 - 3\delta_2}\tilde{v}_2 = 0\\
\end{equation}
\begin{equation}
e^{-\alpha}\tilde{c}_1=\frac{3\tilde{v}_2}{4(1-\delta_2)}\left(1+\delta_2 \pm \left[(1-3\delta_2)^2 -\frac{4(1-\delta_2)(1-3\delta_2)^2}{9\tilde{v}_2^2(1+\epsilon)}\right]^{1/2}\right).
\end{equation}
which are identical to Eqs.(A.39) and (A.45) found at y=0 with $\tilde{c}_1$ replaced by $e^{-\alpha}\tilde{c}_1$ and $\delta_1$ replaced by ($-\delta_2$).  In the stiff potential limit, $\delta_2$ $\to$ 0, this gives $\tilde{v}_1$ = $\tilde{v}_2$ and $e^{-\alpha}$ = 1 which again would not give the desired solution to the hierarchy problem just as in the n=1 case.  

We can also determine the situation for $\delta_i$ small but non-zero.  Table 1 gives some sample values.  Thus a hierarchy is not obtained if we truncate at n=2.  The above results suggest that keeping a finite number of terms in Eqs.(A.10) and (A.11) will not lead to a valid approximation, and it may be that truncating at n=1 does not approximate the rigorous solutions of Eqs.(A.7)-(A.9).
\begin{table}
\begin{tabular}{|r|l|l|l|l|}
\hline
$\epsilon$ & 0.03 & 0.03 & 0.01 & 0.01\\
$\delta_1$ & 0.01 & 0.01 & 0.0001 & 0.0001\\
$\delta_2$ & -0.01 & 0.01 & -0.0001 & 0.0001\\
\hline
$\tilde{v}_1$ & $\pm$.6613 & $\pm$.6613 & $\pm$.6578& $\pm$.6642\\
$\tilde{v}_2$ & $\pm$.6613 & $\pm$ .6542& $\pm$.6578& $\pm$.6640\\
$\tilde{c}_1$ & $\pm$.5152 & $\pm$.5152& $\pm$.5184& $\pm$.5219\\
$e^{-\alpha}$ & 1 & 0.9876 & 1 & 1.0001\\
$e^{-\beta}$ & 1 & 0.6597 & 1 & 1.0101\\ 
\hline
\end{tabular}
\caption{Example of determination of the hierarchy parameter $e^{-\alpha}$ for various choices of $\epsilon$, $\delta_1$, and $\delta_2$.  A valid hierarchy is obtained when $e^{-\beta}$ $\approx$ $10^{-16}$, which requires $\tilde{v_1}/\tilde{v_2}$ $\approx$ 1/3 for $\epsilon$ = .03 and $\tilde{v_1}/\tilde{v_2}$ $\approx$ 2/3 for $\epsilon$ = .01}
\end{table}

\end{document}